\documentstyle[twoside,fleqn,espcrc2,epsf,graphicx]{article}

\title{
Theta vacuum effects on the pseudoscalar condensates and
the $\eta^{\prime}$ meson in 2-dimensional lattice QED
\thanks{Talk presented by
H. Fukaya.}}

\author{\
Hidenori Fukaya
\address[Ox]{Yukawa Institute for Theoretical Physics,Kyoto
University,Kyoto 606-8502,Japan}, Tetsuya Onogi
\addressmark[Ox]\thanks{
Supported by Grant-in-Aid for Scientific research 
from the Ministry of Education, Culture, Sports, Science and 
Technology of Japan (Nos. 13135213, 16028210, 16540243).}}

\begin{document}

\begin{abstract}
We study the chiral condensates and the $\eta^{\prime}$ meson 
correlators of the massive Schwinger model in $\theta\neq 0$ vacuum. 
Our data suggest that 
the pseudoscalar operator does condense in a fixed topological
sector and gives long range correlations 
of the $\eta^{\prime}$ meson.
We find that this is well understood 
from the clustering decomposition and statistical picture.
Our result also indicates that even in $\theta =0$ case, 
the long range correlation of $\eta^{\prime}$ meson 
receives non-zero contributions 
from all the topological sectors and that their cancellation 
is non-trivial and requires accurate measurement of the 
reweighting factors as well as the expectation values.
It is then clear that the fluctuation of the 
``disconnected'' diagram originates from the pseudoscalar 
condensates.
\vspace{0mm}
\end{abstract}

\maketitle
\section{Introduction}
In QCD or the massive Schwinger model 
in $\theta=0$ vacuum
\cite
{Coleman:1975pw,Coleman:1976uz,Smilga:1996pi,Hetrick:1995wq}, 
it is well known that the scalar operator condense while the
pseudoscalar does not;
\begin{eqnarray}
\langle \bar{\psi}\psi \rangle \neq 0, \;\;\;\;\;
\langle \bar{\psi}\gamma_5\psi \rangle = 0
\;\;\;\;\;(\theta=0),
\end{eqnarray}
where the second equation follows from parity symmetry.
However, if we have a non-zero $\theta$ term, which violates parity
symmetry, both of them have non-zero expectation values;
\begin{eqnarray}
\langle \bar{\psi}\psi \rangle \neq 0, \;\;\;\;\;
\langle \bar{\psi}\gamma_5\psi \rangle \neq 0
\;\;\;\;\;(\theta\neq0),
\end{eqnarray}
which indicates that $\eta$ meson
should have a long-range correlation as
\begin{eqnarray}
\lim_{|x|\to \mbox{\scriptsize large}}
\langle \eta^{\dagger}(x)\eta(0)\rangle \propto
\langle \bar{\psi}\gamma_5\psi(x) \rangle
\langle \bar{\psi}\gamma_5\psi(0) \rangle\neq 0.
\end{eqnarray}

We would like to present our numerical results of
the 2-flavor massive Schwinger model with a $\theta\neq 0$ term.
We investigate $\bar{\psi}\gamma_5\psi$ condensates and the 
$\eta^{\prime}$ meson correlators in each topological
sector. It is also found that they are non-trivially related 
each other to reproduce the $\theta$ dependence.
We find that their behavior is well understood by the
intuitive picture based on the clustering decomposition
and the statistical mechanics.

In particular, our study shows that the accurate contributions from
higher topological sectors are essential in order to assure
parity symmetry, which never allows the 
long-range correlation of the $\eta^{\prime}$ meson correlators.
Moreover, it is also shown that the origin of the fluctuations
of disconnected diagram is from pseudoscalar condensates in each
topological sector.

\section{Strategy for simulations}

Our strategy to calculate the $\theta$ vacuum effects is to
separate the integral of the gauge fields into topological sectors;
\begin{eqnarray}\label{eq:exp}
\langle  O \rangle ^{\theta}_{\beta, m} =
\frac{\sum_{Q=-\infty}^{+\infty}e^{iQ \theta}
       \langle  O \rangle^{Q}_{\beta, m}Z^{Q}(\beta, m) }
     { \sum_{Q=-\infty}^{+\infty}e^{iQ \theta}Z^{Q}(\beta, m) },
\end{eqnarray}
where $\beta=1/g^2$ denotes a coupling constant 
and $\langle  O \rangle^{Q}$ and $Z^{Q}$ denote the expectation 
value and the partition function in a fixed topological sector
respectively.

The expectation values with a fixed topological charge;
$\langle  O \rangle^{Q}$ are evaluated
by generating link variables with the following gauge action
\cite{Luscher:1998du};
\begin{eqnarray}
S_{G}&=& \left\{\begin{array}{l}
\displaystyle{\sum_{P}}\frac{(1-{\rm Re}P_{\mu\nu}(x))}
{1-(1-{\rm Re}P_{\mu\nu}(x))/\epsilon} 
\\ \;\;\; \;\;\; \;\;\; \;\;\; \;\;\; \;\;\; \;\;\; \;\;\;\mbox{if admissible}
\\ \infty \;\;\; \;\;\; \;\;\; \;\;\;  \;\;\; \;\;\; \;\;\;\mbox{otherwise},
\end{array}
\right.  \label{eq:admaction2}
\end{eqnarray}
where $P_{\mu\nu}$ denotes a plaquette and $\epsilon$ is a
fixed constant.
This action impose the L\"uscher's bound \cite{Luscher:1998du} 
on the gauge fields, which realize an exact topological charge on the
lattice, that is never changed in each step of the the hybrid Monte Carlo updation.

$Z^Q$ normalized by that of zero topological
sector can be evaluated by decomposing it into three parts;
\begin{eqnarray}
R^Q(\beta,m)\equiv \frac{Z^Q(\beta,m)}{Z^0(\beta,m)}
=\underbrace{e^{-\beta S_{G min}^Q}}_{\mbox{\scriptsize classical solution}}\nonumber\\ 
\;\;\;\;\;\;\times\underbrace{\frac{\int dA^Q_{cl}\det (D + m)^2}
{\int dA^0_{cl}\det (D + m)^2}}_{\mbox{\scriptsize moduli integral}}
\underbrace{e^{\int_{\beta}^{\infty}d\beta^{\prime}\Delta S^Q
(\beta^{\prime},m)}}_{\mbox{\scriptsize fluctuation}},
\end{eqnarray}
where $S_{G min}^Q$ denotes the classical minimum of the gauge action
with topological charge $Q$, $\int dA^Q_{cl}$
denotes the moduli integral, and $\Delta S^Q\equiv 
\langle S_G - S_{G min}^Q\rangle^Q_{\beta^{\prime}}
-\langle S_G\rangle^0_{\beta^{\prime}}$, all of which are numerically 
calculable \cite{Fukaya:2003ph,Fukaya:2004kp}.

We choose the domain-wall fermion action with Pauli-Villars regulators
for sea quarks. The link variables are updated by the hybrid Monte Carlo
algorithm. The parameters are chosen as $\beta=1.0$,
$m=0.1,0.15,0.2,0.25,0.3$. We take $L_x=L_y=16$ and $L_s=6$ lattice
where $L_s$ denotes the size of the extra dimension of domain-wall
fermions. 
50 molecular dynamics steps with a step size $\Delta\tau=0.035$
are performed in one trajectory. Configurations are updated every 10 
trajectories. For each topological sector, around 500 trajectories 
are taken for the thermalization staring from the initial configuration 
which is the classical instanton solution with topological charge $Q$. 
We generate 300 configurations in $-|Q_{max}|\leq Q\leq |Q_{max}|$ sectors for the
measurements and from 1000 to 10000 for the reweighting factors at
various $\beta$, where $|Q_{max}|=4$ at $m=0.1,0.15,0.2$ and 
$|Q_{max}|=5$ at $m=0.25,0.3$.

\section{Results}

\begin{figure}[htp]
\begin{center}
\includegraphics[width=7cm]{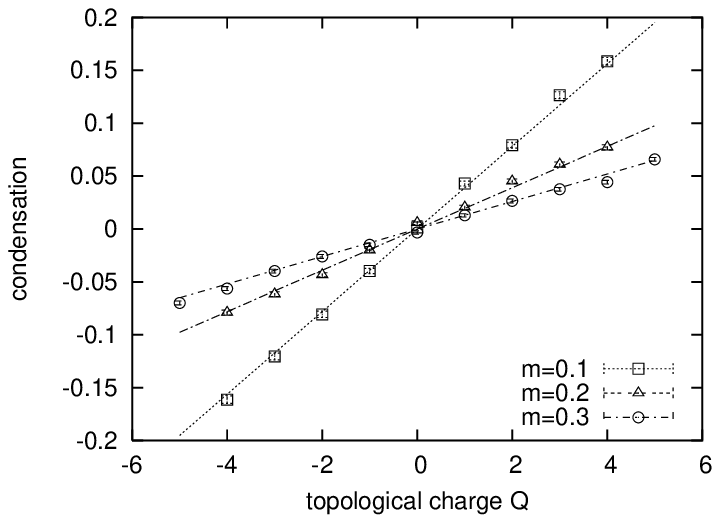}
\vspace{-10mm}
\caption{$-\langle\bar{\psi}\gamma_5\psi\rangle^Q$ 
in each topological sector at $\beta=1.0$.
The lines show the anomaly equation
$-\langle\bar{\psi}\gamma_5\psi\rangle^Q = Q/(mV)$, which 
agrees with our data.}
\label{fig:condaxQ}
\includegraphics[width=7cm]{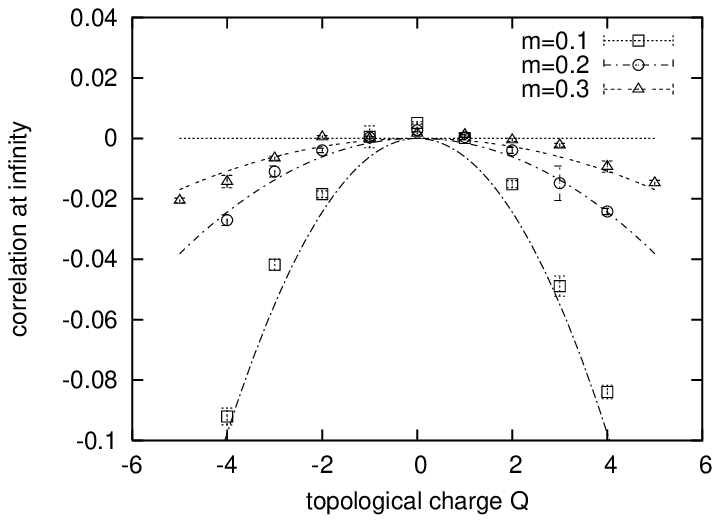}
\vspace{-10mm}
\caption{
The long-range correlations in each topological sector
evaluated by the fit with constants.
The lines are $
\langle \eta^{\prime\dagger}(x)\eta^{\prime}(0)\rangle^Q
\stackrel{|x|\to\mbox{\scriptsize large}}{\to}
 -4 Q^2/(mV)^2$. 
}
\label{fig:longQ}
\includegraphics[width=7cm]{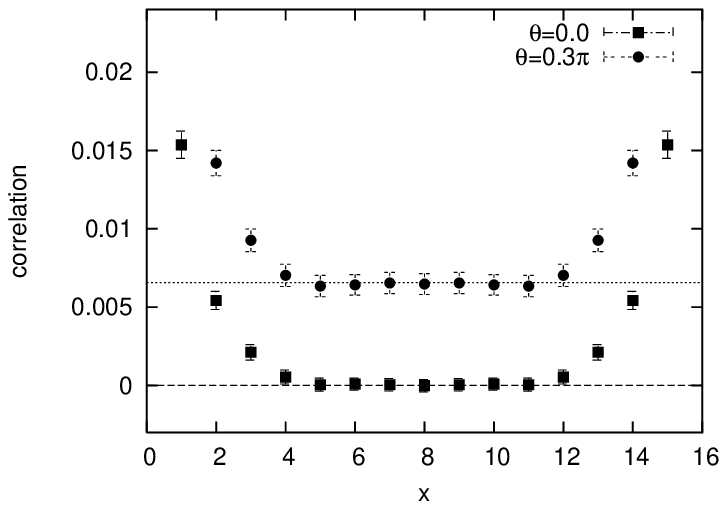}
\vspace{-10mm}
\caption
{
The propagations of the $\eta^{\prime}$ meson at $\theta=0, 0.3\pi$ and
 $m=0.1$ are shown. The lines show the result of the fit.
}
\label{fig:etaproptheta}
\end{center}
\end{figure}

The topological charge dependence of pseudoscalar condensates
is derived by the anomaly equation;
\begin{eqnarray}\label{eq:anom}
-\langle\bar{\psi}\gamma_5\psi\rangle^Q=Q/(mV),
\end{eqnarray}
where $V$ denotes the volume of the torus.
As seen in Fig.~\ref{fig:condaxQ}, our data show a good agreement with 
this equation.

Then the $\eta^{\prime}$ meson correlators should have long-range correlations.
From the clustering decomposition, this can be expressed as
\begin{eqnarray}
\left\langle \eta^{\prime\dagger}(x)\eta^{\prime}(0)\right\rangle^Q
\stackrel{|x|\to\mbox{\scriptsize large}}{\to}
\hspace{1.2in}
\nonumber\\
-\sum_{Q^{\prime}} 
P_{Q,Q^{\prime}}
\langle \sum_f\bar{\psi}_f\gamma_5\psi_f
\rangle^{Q^{\prime}}_B
\langle \sum_f\bar{\psi}_f\gamma_5\psi_f
\rangle^{Q-Q^{\prime}}_A,
\end{eqnarray}
where  $\langle\rangle^{Q^{\prime}}_{A, B}$ means
the expectation value with the topological charge $Q^{\prime}$ 
in the region $A, B$ which denote the half of the large box, where
the pseudoscalar operators reside, respectively 
and $P_{Q,Q^{\prime}}$ denotes the probability 
of the distribution where $Q^{\prime}$ instantons appear in
the box $B$ and $Q-Q^{\prime}$ appear in the box $A$.

In $Q=0$ case, one obtains
\begin{eqnarray}
\left\langle \eta^{\prime\dagger}(x)\eta^{\prime}(0)\right\rangle^0
\stackrel{|x|\to\mbox{\scriptsize large}}{\to}
\hspace{1.2in}
\nonumber\\
-\sum_{Q^{\prime}} 
P_{0,Q^{\prime}}
\langle \sum_f\bar{\psi}_f\gamma_5\psi_f
\rangle^{Q^{\prime}}_B
\langle \sum_f\bar{\psi}_f\gamma_5\psi_f
\rangle^{-Q^{\prime}}_A\nonumber\\
=+\sum_{Q^{\prime}}P_{0,Q^{\prime}}\left(
\langle \sum_f\bar{\psi}_f\gamma_5\psi_f
\rangle^{Q^{\prime}}_B\right)^2>0,
\end{eqnarray}
where we assume $\langle O\rangle_A=\langle O\rangle_B$ and use 
the anti-symmetry;$\langle \bar{\psi}\gamma_5\psi
\rangle^{-Q^{\prime}}=-\langle \bar{\psi}\gamma_5\psi
\rangle^{Q^{\prime}}$ as seen in Fig.~\ref{fig:condaxQ}.
On the other hand, at large $Q$, assuming the distribution $P_{Q,Q^{\prime}}$
to be Gaussian around $Q^{\prime}\sim Q/2$, the correlation can be 
evaluated as follows,
\begin{eqnarray}
\left\langle \eta^{\prime\dagger}(x)\eta^{\prime}(0)\right\rangle^Q
\stackrel{|x|\to\mbox{\scriptsize large}}{\to}
\hspace{1.2in}
\nonumber\\\nonumber\\
-\sqrt{\alpha / \pi}\int dQ^{\prime} 
e^{-\alpha (Q^{\prime}-Q/2)^2}\hspace{1in}\nonumber\\
\times\underbrace{\langle \sum_f\bar{\psi}_f\gamma_5\psi_f
\rangle^{Q^{\prime}}_B}_{\sim 4Q^{\prime}/mV}
\underbrace{\langle \sum_f\bar{\psi}_f\gamma_5\psi_f
\rangle^{Q-Q^{\prime}}_A}_{\sim 4(Q-Q^{\prime})/mV}\nonumber\\
=
-\frac{4}{m^2V^2}\left(Q^2-\frac{1}{\alpha}\right)\sim-\frac{4Q^2}{m^2V^2},
\hspace{.5in}
\end{eqnarray}
where $1/\sqrt{\alpha} << Q/2$ is a numerical constant.
As seen in Fig~\ref{fig:longQ}, it is surprising that 
these very simple arguments describe the data quite well.

$\theta$ dependence of the $\eta^{\prime}$ correlators 
are evaluated by substituting the data into Eq.(\ref{eq:exp}).
Fig.~\ref{fig:etaproptheta} shows the result.
It is obvious that 
there are long-range correlations at $\theta\neq 0$ while
$\theta=0$ case is consistent with zero, which suggests
our reweighting method works well at small $\theta$.

\section{Summary}

We study $Q$ and $\theta$ dependence of the pseudoscalar condensates
and the $\eta^{\prime}$ meson correlators.
We find that pseudoscalar does condense in each topological sector; 
\begin{eqnarray}
-\langle \bar{\psi}\gamma_5\psi \rangle^Q=\frac{Q}{mV}\neq 0,\hspace{1in}
\end{eqnarray}
and there exists a long-range correlation of $\eta^{\prime}$ meson;
\begin{eqnarray}
	\lim_{|x|\to \mbox{\scriptsize large}}
	 \langle\eta^{\prime\dagger}(x)
	 \eta^{\prime}(0)\rangle^{Q=0}>0,\hspace{0.65in}\nonumber\\
	\lim_{|x|\to \mbox{\scriptsize large}}
	 \langle\eta^{\prime\dagger}(x)
	 \eta^{\prime}(0)\rangle^{Q>2}\sim -\frac{4Q^2}{m^2V^2}<0,
\end{eqnarray} 
which are well understood by the clustering properties.

It is also found that each contribution from
different topological sectors plays very important role to
produce non-trivial $\theta$ dependence of these observables.
In particular, the cancellation the long-range correlation of
$\eta^{\prime}$ meson requires accurate measurements of higher
topological sectors. It is also obvious that the fluctuation of
the disconnected diagrams originates from these pseudoscalar 
condensates.

It would be interesting to extend our studies to
4-dimensional QCD.

\end{document}